\documentstyle[12pt,righttag]{amsart}

\newtheorem{prop}{Proposition}[section] 

\theoremstyle{remark}

\numberwithin{equation}{section}
\begin{document}

\title[Quasi-BiHamiltonian Systems and  \dots] {Quasi-BiHamiltonian Systems
and Separability } 

\author[C. Morosi]{C. Morosi \dag}
\address {\dag C. Morosi, Dipartimento di Matematica, Politecnico di Milano,
Piazza  L. Da Vinci 32, I-20133 Milano, Italy}
\email{carmor@@mate.polimi.it}

\author[G. Tondo]{G. Tondo \ddag} 
\address {\ddag G. Tondo, Dipartimento di Scienze Matematiche,  Universit\`a
degli Studi di Trieste,   Piaz.le Europa 1, I-34127  Trieste, Italy.} 
\email{tondo@@univ.trieste.it}

\date{\dag Dipartimento di Matematica, Politecnico di Milano,  Piazza L. Da
Vinci 32, I-20133  Milano, Italy \\ 
\ddag Dipartimento di Scienze Matematiche, Universit\`a degli Studi di Trieste, 
\\  Piaz.le Europa 1, I-34127 Trieste, Italy.} 


\maketitle

\begin{abstract}  Two  quasi--biHamiltonian systems with three and four
degrees of freedom are presented. These systems are shown to be separable 
in terms of Nijenhuis coordinates. Moreover the most general Pfaffian
quasi-biHamiltonian system  with an arbitrary number of degrees of freedom
 is constructed (in terms of Nijenhuis coordinates) and its separability is
proved.
\end{abstract}

\pagebreak

\section {Preliminaries} \label{sec:intro} As it is known, the biHamiltonian
structure is a peculiar  property of  integrable systems, both finite and infinite
dimensional \cite{Magri1, Olver}. We recall  some definitions. Let $M$ be a
differentiable manifold,  $TM$ and $ T^*M$ its tangent and  cotangent bundle
 and $P_0$, $P_1:T^*M\mapsto TM$  two compatible Poisson tensors on
$M$ \cite{Magri1}: a vector field X is said to be biHamiltonian  w.r.t. $P_0$ and 
$P_1$ if there exist two smooth functions $H$ and $F$ such that

\begin{equation}  \label{eq:BHX} X=P_0\, dH=P_1\, dF \ ,
\end{equation}
$d$ denoting the exterior derivative.  Moreover,  if $P_0$ is invertible, the
tensor
$N:=P_1P_0^{-1}$ is a Nijenhuis (or hereditary) tensor; in terms of the 
gradients of the Hamiltonian functions, the biHamiltonian property
 (\ref{eq:BHX}) entails that
$N^*$ (the adjoint map of $N$) maps iteratively $dH$ into closed one--forms, 
so that   $d(N^{*^{i}}dH)=0\ (i=1,2,\ldots)$.  
\par  As a matter of fact, it is in general quite difficult to construct directly a 
biHamiltonian
 structure for a given integrable Hamiltonian vector field; so one can try to
use  some reduction procedure, starting from  a few ``universal'' Poisson
structures defined in an extended phase space.  On the other hand, in the case
of finite--dimensional systems arising as  restricted or stationary flows from
soliton equations \cite{AW,Ton1}, the final result of the reduction procedure
are  some physically interesting dynamical systems (for example the
H\'enon--Heiles system) which,  
 in their natural phase space, satisfy a weaker condition  than the
biHamiltonian one. So,   the  notion of  quasi-biHamiltonian (QBH)  system can
be introduced \cite{ francesi0, francesi1}; it  was applied in 
\cite{francesi2} to dynamical systems with two degrees of freedom. One of the
aims of this paper is just to give explicit examples of QBH systems with more
than two degrees of  freedom.
\par According to \cite{francesi2}, a vector field 
 $X$ is said to be a quasi--biHamiltonian (QBH) vector field  w.r.t. two 
compatible Poisson tensors $P_0$ and $P_1$ if there are three smooth
functions $H,F,\rho$  such  that 

\begin{equation}  \label{eq:QBHX}  X=P_0\, dH= \frac{1}{\rho} P_1\, dF 
\end{equation}  ($\rho$ playing the role of an integrating factor). From this
equation it follows   that $F$ is an integral of motion for
$X$, in involution with $H$, so that a QBH vector field with two degrees of 
freedom is Liouville-integrable. Of course, Eq.(\ref{eq:QBHX}) can be studied
for an  arbitrary number $n$ of degrees of freedom, but the knowledge of $F$
and $H$ is no  more sufficient to assure the integrability of $X$ for $n >2$. In
this case, the search for the  integrability can be pursued using a sufficient
criterion, which was recently introduced by one  of the present authors (G.T.).
Indeed,  one can show that 

\begin{prop} \label{pr:nostroQBH} \cite{Ton1}
 Let  $M$ be a  $2n$ dimensional symplectic manifold equipped with  an
invertible Poisson tensor  
$P_0$, and  $X$  be a Hamiltonian vector field
 with Hamiltonian $H$: $X=P_0 \, dH$. Let there exist  a tensor  
$N:TM\rightarrow TM$  such that the tensor  
$P_1:T^*M\rightarrow TM$  defined by  $P_1:=N P_0$ is   skew--symmetric.
Denote by $ X_i:=N^{i-1} X$  and  
 $ \alpha_{i}:=N^{*^{i-1}}dH \  (i=1,2,\ldots)$ the vector fields and the 
one--forms  obtained by the iterated action of $N$ and $N^*$.  
\par 
 If  there exist $(n-1)$ independent functions $H_i\  (i=2,\ldots ,n)$ and
$(n(n+1)/2 -1)$ functions $\rho_{ij} \  (i=2,\ldots,n; 1\leq j \leq i)$ with
$\rho_{11}=1$ and $\rho_{ii}\neq 0\ (i=2,\ldots,n)$,   such that the 1-forms 
$\alpha_{i}$  can be written as  
$\alpha_i =\sum_{j=1}^i \rho_{ij}\, dH_j\  (i=1,2,\ldots,n)$, then: 
\par
\par\noindent
 {\bf i)}
 the vector fields $X_i$ satisfy the recursion relations  
$ X_{i+1}=P_0\,\alpha _{i+1}=P_1\,\alpha _{i} \  (i=1,\ldots, n-1)$;
\par
\par\noindent   {\bf ii)}
  the functions $H_i$ are in involution with respect to the Poisson bracket 
defined by
$P_0$ and they are constants of motion for each field $X_k$ $(k=1,\ldots, n)$;
\par\noindent
 {\bf iii)}  the Hamiltonian system corresponding to the vector field
$X$  is Liouville--integrable.  Moreover, if $P_1$  is a Poisson tensor, then also
$X_2$  is an  integrable Hamiltonian vector field and the functions $H_i$ are in
involution also with respect to the Poisson bracket defined   by $P_1$.
\end{prop} This result is applied in the next section of this paper, where we
consider two  H\'enon--Heiles  type systems with three and four degrees of
freedom.
\par To fix the notations, on any open set of a  $2n$ dimensional symplectic
manifold 
$M$, let  (${\boldsymbol q}=(q_1, \ldots, q_n)$; 
${\boldsymbol p}=(p_1,\ldots,p_n)$) be a set of  canonical coordinates and 
$P_0$ the Poisson tensor
$P_0=
\begin{bmatrix}  0&{\boldsymbol I}\\\ -{\boldsymbol I}&0 
\end{bmatrix} 
$  (${\boldsymbol I}$ denoting the $n \times n$ identity matrix). Let $P_1$ be
a compatible Poisson tensor w.r.t. $P_0$, such that the Nijenhuis tensor 
$N:=P_1P_0^{-1}$ is maximal, i.e., it has $n$ distinct eigenvalues
${\boldsymbol \lambda}=(\lambda_1,\ldots, \lambda_n)$. As it is known 
\cite{TMagri}, in a neighborhood of a regular point, where the eigenvalues
${\boldsymbol \lambda}$ are independent,  one can construct a canonical 
transformation
$({\boldsymbol q};{\boldsymbol p})\mapsto ({\boldsymbol \lambda}; 
{\boldsymbol\mu)}$ ($({\boldsymbol\lambda};{\boldsymbol\mu)} $ referred
to as  Nijenhuis  coordinates) such that
$P_1$ and $N$ take the Darboux form 

\begin{equation} \label{eq:DarbP1N} P_1 = 
\begin{bmatrix}  0&\Lambda\\\ -\Lambda &0 
\end{bmatrix} \ , \qquad  N=
\begin{bmatrix} 
\Lambda&0 \\\ 0&\Lambda  
\end{bmatrix}  \qquad  
\left(\Lambda:=diag(\lambda_1, \ldots, \lambda_n)\right)\ .
\end{equation}  A QBH vector field is said to be Pfaffian
\cite{francesi2} if the integrating factor  $\rho$ in Eq.(\ref{eq:QBHX}) is the 
product of the eigenvalues of $N$, i.e.,  

\begin{equation}   \label{eq:Pfaff} 
\rho=\Pi_{i=1}^n \lambda_i \ . 
\end{equation}  Working in this setting,

\begin{itemize}
\item  in Sect.2 we present two  H\'enon--Heiles type systems with three and
four  degrees of freedom, which are Pfaffian QBH systems; passing to a set of 
Nijenhuis coordinates, we show that the  Hamilton-Jacobi equations  for these
systems are separable; 
\item  in Sect.3 we obtain the general solution of Eq.(\ref{eq:QBHX}) for a
Pfaffian QBH vector field with an arbitrary
 number of degrees of freedom; the Hamiltonian
$H$ and the function $F$ contain  $n$ arbitrary smooth functions $f_i$, each 
one of them depending on a single pair  $(\lambda_i; \mu_i)$ of Nijenhuis
coordinates.  Finally, we prove  that the Hamiltonian $H$  is separable. 
\end{itemize}

\section{Two H\'enon--Heiles type systems with three and four degrees of
freedom}  In this section we present two separable  QBH   systems with three
and  four degrees of freedom; they  belong to a family of integrable flows
obtained in
\cite{Ton1} as stationary flows  of the Korteweg--de Vries hierarchy \cite{Dic}.
This family contains the classical H\'enon--Heiles system as its second
member, so the higher members can be considered as multi-dimensional
extensions of  H\'enon--Heiles. 
\par
 The third member of this family,  which is a stationary reduction of the
seventh order KdV flow, is defined in a six dimensional  phase space (with
coordinates  ${\boldsymbol q}=(q_1,q_2,q_3)$,  
${\boldsymbol p}=(p_1, p_2,p_3)$) by the   Hamiltonian vector field   $X=P_0\,
dH$, with Hamiltonian function

\begin{equation}\label{H1}  H=\frac{1}{2}(2p_1p_2+p_3^2)-\frac{5}{8}q_1^4
+\frac{5}{2}q_1^2q_2+\frac{q_1q_3^2}{2}-\frac{q_2^2}{2} .
\end{equation} 
First of all, we can show that the vector field   $X$ is
Liouville-integrable. Indeed, if one introduces the functions

\begin{equation} \label{H23}
\begin{split} H_1&=H \ , \\ 
H_{2}&=\frac{p_1^2}{2}+p_1 p_2 q_1+p_3^2
q_1-p_2^2 q_2-p_2 p_3 q_3-\frac{q_1^5}{2} -\frac{q_1^2 q_3^2}{4}+\frac{q_2
q_3^2}{2}+2q_1 q_2^2\ , \\ 
 H_{3}&=\frac{p_3^2q_1^2}{2}+p_3^2 q_2-p_1 p_3 q_3-p_2 p_3 q_1 q_3+
\frac{p_2^2 q_3^2}{2}+\frac{q_1^3 q_3^2}{2}-q_1 q_2 q_3^2-\frac{q_3^4}{8}  \ ,
 \end{split}
\end{equation} 
 $X$ satisfies the assumptions of Prop.\ref{pr:nostroQBH};   the  tensor $P_1$ is
given by

\begin{equation} \label{eq:P1HH3D}  
P_1 = 
\begin{bmatrix}  
0&A\\\ -A^T&B 
\end{bmatrix} \ , 
\quad  A=-
\begin{bmatrix}  
q_1&-1&0 \\\ 
2q_2&q_1&q_3 \\\ 
q_3&0&0
\end{bmatrix}  \ , 
\quad  B=
\begin{bmatrix}  
0&-p_2&-p_3 \\\ 
p_2&0&0 \\\  
p_3&0&0 
\end{bmatrix}  \ ,
\end{equation} and the functions $\rho_{ij}$ are:
$\rho_{11}=\rho_{22}=\rho_{33}=1$,
$\rho_{21}=\rho_{32}=-2q_1$,
$\rho_{31}=(3q_1^2-2q_2)$. 
\par Furthermore one easily verifies that $P_1$ is a Poisson tensor,
compatible with $P_0$ (so that
$N=P_1P_0^{-1}$ is a Nijenhuis tensor). One can show that $X$ is a QBH vector
field;   
 in fact Eq.(\ref{eq:QBHX}) is verified with $\rho$ and $F$ given by 
 $\rho=q_3^2$ and $F=H_3$. 
\par 
 At last, let us show the separability of this system in terms of Nijenhuis
coordinates.  In this case the construction of a canonical map  
$\Phi: ({\boldsymbol \lambda}; {\boldsymbol \mu}) \mapsto ({\boldsymbol 
q};{\boldsymbol  p})$ between a set of Nijenhuis coordinates $({\boldsymbol 
\lambda}; {\boldsymbol \mu})$ and the  coordinates $({\boldsymbol
q};{\boldsymbol p})$ is quite simple. We observe  that the matrix $A$ in
Eq.(\ref{eq:P1HH3D}) depends only on the coordinates ${\boldsymbol  q}$, so
also the eigenvalues  ${\boldsymbol  \lambda}$ depend only on
${\boldsymbol  q}$: 
 $q_k=f_k( {\boldsymbol  \lambda})$. Then  we introduce the  generating
function $S=\sum_{k=1}^3 p_k f_k( {\boldsymbol  \lambda})$ and we get

\begin{equation}  \label{eq:Nmap3D} 
\begin{split} 
 q_1&=-\frac{1}{2}(\lambda_1+\lambda_2+\lambda_3) \ , \\ 
q_2&=-\frac{1}{8}(\lambda_1+\lambda_2+\lambda_3)^2
+\frac{1}{2}(\lambda_1\lambda_2+\lambda_1\lambda_3+\lambda_2\lambda_3)
\ , \\
 q_3&= (\lambda_1\lambda_2\lambda_3)^{1/2} \ ,  \\ 
p_1&=
\frac{\lambda_1\mu_1}{\lambda_{12}\lambda_{13}}(-\lambda_1+\lambda_2+
\lambda_3)
+
\frac{\lambda_2\mu_2}{\lambda_{21}\lambda_{23}}(\lambda_1-\lambda_2+
\lambda_3)
+ 
\frac{\lambda_3\mu_3}{\lambda_{31}\lambda_{32}}(\lambda_1+\lambda_2-
\lambda_3),
\\  
p_2&=-2(\frac{\lambda_1\mu_1}{\lambda_{12}\lambda_{13}}+
\frac{\lambda_2\mu_2}{\lambda_{21}\lambda_{23}}+ 
\frac{\lambda_3\mu_3}{\lambda_{31}\lambda_{32}}) \ , \\
  p_3 &= 2(\lambda_1\lambda_2\lambda_3)^{1/2} 
 ( \frac{\mu_1}{\lambda_{12}\lambda_{13}}+ 
\frac{\mu_2}{\lambda_{21}\lambda_{23}} +
\frac{\mu_3}{\lambda_{31}\lambda_{32}})  \ ,
\end{split}
\end{equation} 
where we  put, for brevity,
$\lambda_{ij}:=\lambda_i-\lambda_j$.  Since
$\rho=q_3^2=\lambda_1\lambda_2\lambda_3$, we are faced  with a Pfaffian
system. Written in the above mentioned Nijenhuis coordinates, the
Hamiltonian  function $H$ given by Eq.(\ref{H1}) takes the form

\begin{equation}  \label{eq:H1N}  
H=\frac{\lambda_1 (16
\mu_1^2-\lambda_1^5)}{8\lambda_{12}\lambda_{13}}+
    \frac{\lambda_2 (16 \mu_2^2-\lambda_2^5)}{8\lambda_{21}\lambda_{23}}
+
     \frac{\lambda_3 (16
\mu_3^2-\lambda_3^5)}{8\lambda_{31}\lambda_{32}} \ .
\end{equation}
 It is easy to show that  the Hamilton-Jacobi equation 
$H({\boldsymbol  \lambda}, \frac{\partial W}{\partial {\boldsymbol
\lambda}})=h$ 
 is separable and has the complete integral
$W=\sum_{i=1}^3 W_i(\lambda_i; c_0,c_1,c_2)$, with
$W_1$, $W_2$ and
$W_3$ solutions of  the following equations 

\begin{equation} 
\frac{dW_i}{d\lambda_i}=
 \left(
\frac{1}{16\lambda_i}
(\lambda_i^6+c_2\lambda_i^2+c_1\lambda_i+c_0)\right)^{1/2} 
 \quad    c_2=8h \ , \qquad (i=1,2, 3) \ .
\end{equation}
\par
\par

Our second example is a  H\'enon--Heiles system with four degrees of
freedom. It can be constructed as  a stationary reduction of the  ninth order
KdV flow  \cite{Ton2}.  Its phase space is eight dimensional, and the
Hamiltonian is

\begin{equation}     \label{eq:HH4D}
 H=\frac{1}{2}(p_4^2+2p_1 p_3 +p_2^2) + 
\frac{3}{4}q_1^5-\frac{5}{2}q_1 ^3q_2+2q_1 q_2^2 + \frac{5}{2}q_1^2q_3 +
\frac{q_1q_4^2}{2}-q_2q_3  \ .
\end{equation}
 Also in this case,  the vector field $X=P_0\,dH$ is Liouville-integrable. Indeed,
let us consider the functions 

\begin{equation} \label{eq:Hj3D}
\begin{split}
H_1&=H \ , \\
H_{2}&=p_1p_2+p_2^2q_1+p_1p_3q_1+p_4^2q_1-p_2p_3q_2
-p_3^2q_3-p_3p_4q_4\\ 
 &+\frac{5}{8}q_1^6- \frac{5}{4}q_1^4q_2- 
q_1^2q_2^2-\frac{q_1^2q_4^2}{4}+q_2^3+\frac{q_2q_4^2}{2}+3q_1q_2q_3-
\frac{1}{2}q_3^2 \ , \\  
 H_{3}&=\frac{1}{2}p_2^2q_1^2+\frac{1}{2}p_4^2q_1^2+
\frac{1}{2}p_3^2q_2^2+p_2p_3q_1q_2+\frac{1}{2}p_3^2q_4^2-p_3p_4q_1q_4 
\\ 
&-2p_2p_3q_3+p_4^2q_2+p_1p_3q_2+p_1p_2q_1-p_2p_4q_4+\frac{1}{2}p_1^2
\\
&+\frac{5}{4}q_1^5q_2-3q_1^3q_2^2+\frac{1}{2}q_1^3q_4^2+\frac{5}{4}q_1^4q_3+
q_1q_2^3-q_1^2q_2q_3-\frac{1}{2}q_1q_2q_4^2  \\ 
&+\frac{1}{2}q_3q_4^2+q_2^2q_3+2q_1q_3^2  \ ,\\
H_{4}&=-p_2p_4q_1q_4-p_3p_4q_2q_4+p_2p_3q_4^2+p_4^2q_1q_2+
p_4^2q_3-p_1p_4q_4\\
&-\frac{5}{8}q_1^4q_4^2+\frac{3}{2}q_1^2q_2q_4^2-
\frac{1}{2}q_2^2q_4^2-q_1q_3q_4^2-
\frac{1}{8}q_4^4   \ ,  
\end{split}
\end{equation}  and the  tensor
$P_1=\begin{bmatrix}
 0&A \\\ 
 -A^T &B
\end{bmatrix} 
$,   with  the matrices $A$ and $B$ given by 

\begin{equation}  \label{eq:P1HH4D}  
A=- \left[
\begin{array}{cccc}  
 q_1&-1&0&0\\\    
q_2&0&-1&0 \\\ 
 2q_3&q_2&q_1&q_4 
\\\   q_4&0&0&0
\end{array} 
\right] \ , \qquad  
B= \left[
\begin{array} {cccc}    
0&-p_2&-p_3&-p_4\\\ 
 p_2&0&0&0 \\\ 
 p_3&0&0&0
\\\  p_4&0&0&0
\end{array} 
\right] \ . 
\end{equation} Then $X$ verifies the assumptions of Prop.\ref{pr:nostroQBH}
with the following choices for the functions $\rho_{ij}$:
$\rho_{11}=\rho_{22}=\rho_{33}=\rho_{44}=1$,
$\rho_{21}=\rho_{32}=\rho_{43}=-2q_1$,
$\rho_{31}=\rho_{42}=(3q_1^2-2q_2)$, $\rho_{41}=(-4q_1^3+6q_1q_2-2q_3)$. 
\par Moreover, $P_1$ is a Poisson tensor, compatible with $P_0$ (so that
$N=P_1P_0^{-1}$ is a Nijenhuis tensor). The Hamiltonian vector field 
$X$ is a QBH vector field since it satisfies the equation $X=P_1 dF/\rho$, with
$\rho=-q_4^2$, $F=-H_4$.
\par
 At last, let us consider the map between the coordinates  
 $({\boldsymbol q};{\boldsymbol p})$ and the Nijenhuis coordinates
$({\boldsymbol  \lambda}; {\boldsymbol \mu})$. Since also in this case the
matrix $A$ in Eq.(\ref{eq:P1HH4D}) depends only on ${\boldsymbol q}$, we
proceed as in the previous example. The result is

\begin{equation} \label{eq:NmapHH4D}
\begin{split} 
 &\lambda_1+\lambda_2+\lambda_3+\lambda_4=-2q_1  \ , \\ 
&\lambda_1\lambda_2+\lambda_1\lambda_3+\lambda_1\lambda_4+
\lambda_2\lambda_3+
\lambda_2\lambda_4+\lambda_3\lambda_4=q_1^2+2q_2 \ ,  \\
&\lambda_1\lambda_2\lambda_3+\lambda_1\lambda_2\lambda_4+
\lambda_2\lambda_3\lambda_4=-2(q_1q_2+q_3)  \ , \\
&\lambda_1\lambda_2\lambda_3\lambda_4=-q_4^2  \ , \\
\mu_1&= -\frac{p_1}{2}-\frac{p_4}{2}\frac{\lambda_2\lambda_3\lambda_4}
{(-\lambda_1\lambda_2\lambda_3\lambda_4)^{1/2}}+  
\frac{p_2}{4}(-\lambda_1+\lambda_2+\lambda_3+\lambda_4)  \\
&+\frac{p_3}{16}(-3\lambda_1^2+2\lambda_1\lambda_2+\lambda_2^2+
2\lambda_1\lambda_3-
2\lambda_2\lambda_3+\lambda_3^2+2\lambda_1\lambda_4-
2\lambda_2\lambda_4-
2\lambda_3\lambda_4+\lambda_4^2)  \ , \\ 
\mu_2&= -\frac{p_1}{2}-\frac{p_4}{2}\frac{\lambda_1\lambda_3\lambda_4}
{(-\lambda_1\lambda_2\lambda_3\lambda_4)^{1/2}}+ 
\frac{p_2}{4}(\lambda_1-\lambda_2+\lambda_3+\lambda_4) \\
&+\frac{p_3}{16}(\lambda_1^2+2\lambda_1\lambda_2-3\lambda_2^2-
2\lambda_1\lambda_3+
2\lambda_2\lambda_3+\lambda_3^2-2\lambda_1\lambda_4+
2\lambda_2\lambda_4-
2\lambda_3\lambda_4+\lambda_4^2)  \ , \\ 
\mu_3&= -\frac{p_1}{2}-\frac{p_4}{2}\frac{\lambda_1\lambda_2\lambda_4}
{(-\lambda_1\lambda_2\lambda_3\lambda_4)^{1/2}}+ 
\frac{p_2}{4}(\lambda_1+\lambda_2-\lambda_3+\lambda_4) \\ 
&+\frac{p_3}{16}(\lambda_1^2-2\lambda_1\lambda_2+\lambda_2^2+
2\lambda_1\lambda_3+
2\lambda_2\lambda_3-3\lambda_3^2-2\lambda_1\lambda_4-
2\lambda_2\lambda_4+
2\lambda_3\lambda_4+\lambda_4^2)  \ ,\\ 
\mu_4&=  -\frac{p_1}{2}-\frac{p_4}{2}\frac{\lambda_1\lambda_2\lambda_3}
{(-\lambda_1\lambda_2\lambda_3\lambda_4)^{1/2}}+ 
\frac{p_2}{4}(\lambda_1+\lambda_2+\lambda_3-\lambda_4) \\ 
&\smash{+\frac{p_3}{16}(\lambda_1^2-2\lambda_1\lambda_2+
\lambda_2^2-2\lambda_1\lambda_3-
2\lambda_2\lambda_3+\lambda_3^2+2\lambda_1\lambda_4+
2\lambda_2\lambda_4+
2\lambda_3\lambda_4-3\lambda_4^2)\ .}
 \end{split}
\end{equation}  By solving this system with respect to $({\boldsymbol 
q};{\boldsymbol  p})$ one can recover  the canonical map $\Phi:
({\boldsymbol  \lambda}; {\boldsymbol  \mu}) \mapsto ({\boldsymbol 
q};{\boldsymbol  p})$ which allows one to write    the Hamiltonian function
$H$ given by Eq.(\ref{eq:HH4D}) in terms of  Nijenhuis coordinates;   it reads 

\begin{equation} \label{eq:H13DN}  
H=\frac{\lambda_1 (16
\mu_1^2-\lambda_1^7)}{8\lambda_{12}\lambda_{13}\lambda_{14}}+
    \frac{\lambda_2 (16
\mu_2^2-\lambda_2^7)}{8\lambda_{21}\lambda_{23}\lambda_{24}} +
     \frac{\lambda_3 (16
\mu_3^2-\lambda_3^7)}{8\lambda_{31}\lambda_{32}\lambda_{34}}+
     \frac{\lambda_4 (16
\mu_4^2-\lambda_4^7)}{8\lambda_{41}\lambda_{42}\lambda_{43}} .
\end{equation}  
Let us remark that also in this case the system is Pfaffian,
since
$\rho=-q_4^2=\lambda_1\lambda_2\lambda_3\lambda_4$. At last, one
proves that the  Hamilton--Jacobi equation
$H({\boldsymbol \lambda};
\frac{\partial W}{\partial {\boldsymbol \lambda}})=h$  is separable and has
the complete integral 
$W=\sum_{i=1}^4W_i( \lambda_i; c_0,c_1,c_2,c_3)$, with $W_1$, $W_2$,
$W_3$ and
$W_4$ solutions of the following equations 

\begin{equation} 
\makebox [0pt]{$\displaystyle \frac{dW_i}{d\lambda_i}$}\quad=
 \left( \frac{1}{16\lambda_i}
\left( \lambda_i^8+c_3\lambda_i^3+c_2\lambda_i^2+c_1\lambda_i+c_0 \right)
\right) ^{1/2} 
 \    c_3=8h  \quad  (i=1,2,3,4).  
\end{equation}
\par

\section{ Quasi-biHamiltonian systems with $n$ degrees of freedom}
\label{sec:nQBHS}  Let us consider a  $2n$ dimensional symplectic manifold
$M$, a Poisson tensor $P_1$ compatible with  $P_0$, and let us assume to
have introduced a set of Nijenhuis coordinates $(\boldsymbol{\lambda} ;
{\boldsymbol \mu})$, so that $P_1$ takes the Darboux form
(\ref{eq:DarbP1N}). We search for the general solution of the QBH equation
(\ref{eq:QBHX}) in the Pfaffian case (i.e., with $\rho$  defined by
Eq.(\ref{eq:Pfaff})). 

\begin{prop} \label{pr:nQBHS}  In the Pfaffian case, the  general solution of
the equation $P_0\, dH=P_1\, dF/\rho$ is given by 
\begin{equation} \label{eq:HFnQBHS}  H=\sum_{i=1}^n
\frac{1}{\Delta_i}f_i(\lambda_i;\mu_i) \ , \qquad  F=\sum_{i=1}^n
\frac{\rho_i}{\Delta_i}f_i(\lambda_i;\mu_i)  \ , 
\end{equation}  where  $(\boldsymbol{\lambda} ; {\boldsymbol \mu})$ are 
Nijenhuis coordinates, 
 $\Delta_i:=\Pi_{j\neq i}\lambda_{ij}$
$(\lambda_{ij}:=\lambda_i-\lambda_j)$, $\rho_i:=\rho/\lambda_i$ and the
$n$ functions $f_i(\lambda_i;\mu_i)$ (each one of them depending on one pair
of coordinates) are arbitrary smooth functions. 
\end{prop}
\begin{pf}   Eq.(\ref{eq:QBHX}) corresponds to the two sets of equations 

\begin{equation} \label{eq:Hmu=Fmu} 
\frac{\partial H}{\partial \mu_i}= \frac{\lambda_i}{\rho}\frac{\partial
F}{\partial \mu_i}
\qquad\qquad   (i=1,2,\ldots,n) \ ,
\end{equation} 
\begin{equation} \label{eq:Hlambda=Flambda} 
\frac{\partial H}{\partial \lambda_i}= \frac{\lambda_i}{\rho}\frac{\partial
F}{\partial
\lambda_i}   \qquad \qquad  (i=1,2,\ldots,n)  \ .
\end{equation}  The general solution of the first set is 

\begin{equation}  \label{eq:solHFmu}  
H=\frac{1}{\rho}\sum_{i=1}^n
\lambda_i\, G_i({\boldsymbol \lambda};\mu_i)+K({\boldsymbol
\lambda}) \ ,
\quad F=\sum_{i=1}^n  G_i({\boldsymbol \lambda};\mu_i) \ , 
\end{equation}  where the functions $G_i=G_i({\boldsymbol \lambda};\mu_i)$
and $K=K({\boldsymbol
\lambda})$ are arbitrary. Indeed, the solution of the first equation
(\ref{eq:Hmu=Fmu}), for
$i=1$, is
$H=\frac{\lambda_1}{\rho}F({\boldsymbol \lambda};{\boldsymbol \mu})+
\phi_1({\boldsymbol
\lambda};
\mu_2,\ldots,\mu_n)$, with
$\phi_1$ arbitrary; on account of this result, the equation (\ref{eq:Hmu=Fmu})
for $i=2$ has the solution 

\begin{equation} 
\begin{split} 
 H&=\frac{\lambda_1}{\rho} G_1({\boldsymbol \lambda};\mu_1)+
\frac{\lambda_2}{\rho} \psi_1({\boldsymbol \lambda};\mu_2,\ldots,\mu_n)+
 \phi_2({\boldsymbol \lambda}; \mu_3,\ldots,\mu_n) \ , \\  F&=
G_1({\boldsymbol \lambda};\mu_1)+\psi_1({\boldsymbol
\lambda};\mu_2,\ldots,\mu_n)
\ , 
\end{split} 
\end{equation}  with $ \psi_1$ and $  \phi_2$ arbitrary. Iterating this
procedure for $i=3,\ldots, n$ one easily obtains the solution
(\ref{eq:solHFmu}).  Let us insert this solution into
Eq.s(\ref{eq:Hlambda=Flambda}), putting into evidence the dependence on
${\boldsymbol \mu}$; we conclude that $K({\boldsymbol
\lambda})$ has to be a constant function (which can be taken equal to zero
with no loss of generality) and that  Eq.s(\ref{eq:Hlambda=Flambda}) can be
written as 

\begin{equation} \label{eq: ?} 
\frac{\partial}{\partial \lambda_i}\left(\sum_{j=1}^n \lambda_{ij}G_j\right)=
\frac{1}{\lambda_i}\left(\sum_{j=1}^n \lambda_{ij}G_j\right)  \qquad
(i=1,2,\ldots,n) \ .
\end{equation}  By integrating these equations $(i=1,2,\ldots,n)$ and taking
into account the dependence on 
${\boldsymbol \mu}$, we easily obtain that 

\begin{equation}  G_i({\boldsymbol \lambda};{\boldsymbol
\mu})=\frac{\rho_i}{\Delta_i}f_i(\lambda_i;\mu_i)
\qquad (i=1,2,\ldots,n) \ ,
\end{equation}  where each $f_i$ is an arbitrary function depending only on
the pair of variables
$(\lambda_i;\mu_i)$.
\end{pf} Of course, the vector field $X=P_0\, dH$ is a QBH vector field in $2n$
dimensions. 
\par At last,  on account of the above result, we can also prove that the
Hamiltonian $H$ and the function $F$  are separable:

\begin{prop} \label{pr:QBHSsep} The Hamiltonian  $H$ and the function $F$,
written in terms of the Nijenhuis coordinates
$({\boldsymbol
\lambda};{\boldsymbol \mu})$ in the form (\ref{eq:HFnQBHS}), are separable
for each $n$-ple of functions
$f_i(\lambda_i;\mu_i)$. 
\end{prop}
\begin{pf} The Hamilton-Jacobi equation for $H$ is separable iff $H$ verifies
the Levi-Civita conditions 
$L_{ij}(H)=0$ $(i,j=1, \ldots, n; i \neq j)$ where \cite{LC}

\begin{equation} \label{eq:LCc} 
\makebox [0pt]{$ L_{ij}(H)\qquad$}=\frac{\partial H}{\partial
\lambda_i}\frac{\partial H}{\partial\lambda_j}
\frac{\partial^2 H}{\partial \mu_i \partial \mu_j}+ 
\frac{\partial H}{\partial \mu_i}\frac{\partial H}{\partial\mu_j}
\frac{\partial^2 H}{\partial \lambda_i \partial \lambda_j}- 
\frac{\partial H}{\partial \lambda_i}\frac{\partial H}{\partial\mu_j}
\frac{\partial^2 H}{\partial \mu_i \partial \lambda_j}- 
 \frac{\partial H}{\partial \lambda_j}\frac{\partial H}{\partial\mu_i}
\frac{\partial^2 H}{\partial \mu_j \partial \lambda_i}.
\end{equation} 
 In our case, it is
$\partial^2H/ \partial\mu_i\partial\mu_j=0$ and 

\begin{equation} \label{eq:derdelta}
\frac{\partial \Delta_j}{\partial \lambda_j}=  
\Delta_j \sum_{\alpha \neq j}\lambda_{j\alpha}^{-1}\ , \qquad 
\frac{\partial \Delta_j}{\partial \lambda_\beta}=  -\Delta_j\,
\lambda_{j\beta}^{-1} \quad (\beta \neq j) \ . 
\end{equation}  It may be useful to decompose  $L_{ij}(H)$  as
$L_{ij}(H)=M_{ij}(H)+N_{ij}(H)$, where
$M_{ij}(H)$ depends linearly on the functions $f_i$, and $N_{ij}(H)$ depends
on the derivatives
$\partial f_i/\partial \lambda_i$ but not on $f_i$. Using  Eq.(\ref{eq:derdelta})
one directly verifies that $M_{ij}(H)=0$ and $N_{ij}(H)=0$. Similarly, one can
show that the Levi--Civita conditions (\ref{eq:LCc}) are fulfilled also by the
function $F$ given in (\ref{eq:HFnQBHS}).
\end{pf}

\vspace{1truecm} {\bf Acknowledgments.} 

We thank an anonymous referee for useful remarks and for pointing out to us 
reference \cite{francesi0}.
\par This work has been partially supported by the GNFM of the Italian CNR
and by the project  "Metodi Geometrici e probabilistici in Fisica Matematica" of
the Italian MURST.

\end{document}